\documentstyle[11pt,paspconf, epsf]{article}

\begin{document}
\title{Gravitational Microlensing: Past, Present and Future}

\author{Shude Mao}
\affil{Max-Planck-Institute f\"ur Astrophysik,
        Karl-Schwarzschild-Strasse 1, 85740 Garching, Germany}

\begin{abstract}
Ongoing microlensing surveys have already yielded more than five
hundred microlensing events, most of which have been identified
in real-time. In this review I present
the basic theory and observational status of these surveys.
I highlight the discoveries made so far -- these
include constraints on the dark matter content of the Galaxy,
the structure and mass function of stars in the Galactic bulge and
applications on stellar atmospheres and populations.
I also briefly discuss the scientific
returns (such as planet detection)
from real-time photometric and spectroscopic follow-up surveys.
With improved software
algorithms and more powerful instrumentations, the number of microlensing
events will be greatly increased in the future; the Space Interferometry
Mission, on the other hand, provides the possibility to break the
lens mass and distance degeneracy. I conclude
that microlensing is developing into an exciting technique with
diverse applications.
\end{abstract}

\section{Introduction}

While we know reasonably well the light distribution in our Milky Way,
the matter content in the Galaxy is not as well understood. From
the rotation curve of the Galaxy, it appears that there is large
amount of dark matter in the outer part of the Galaxy. It is still highly
controversial just in what form the dark matter is: candidates ranging
from elementary particles (e.g., Spergel 1997) to massive astrophysical
candidates have all been proposed.
The latter class of candidates with mass ranging 
from $10^{-7} M_\odot$ to $10^6 M_\odot$ can be probed using
gravitational microlensing.
The principle of this method is based on general relativity; on
the Galactic scale, microlensing involves essentially only
simple Euclidean geometry. The goal of this review
is to present a broad overview of this field and 
point out references where more details can be found.
The structure of the review is as follows. 
In section 2, I present a brief history of microlensing.
The basic theoretical concepts and 
current observational status 
are then presented in \S 3 and 4, respectively. 
In section 5, I discuss the scientific highlights from microlensing
surveys.
And finally in section 6, I
point out the future directions. For other
reviews, see e.g., Paczy\'nski (1996b), Gould (1996), and
Roulet \& Mollerach (1997).
For a review on microlensing
at cosmological distances, see Wambsganss (1999, this volume). For
further details, links to ongoing microlensing surveys and explanations
of acronyms,
see {\tt http://www.mpa-garching.mpg.de/\~\/smao/microlens.html}.

\section{Gravitational Microlensing: A Brief History}

Gravitational (micro)lensing has a long history. It involves many famous
astronomers in the history; for an excellent review, see the
contribution by V. Trimble in this volume. The theory of
(micro)lensing by a point mass was worked out in the framework of general
relativity by Einstein as early as 1912, but was only re-derived and published
in Einstein (1936). The theory was expanded in more detail by Liebes 
(1964)
and Refsdal (1964). After the
discovery of the first lens 0957+561, people's attention turned to
microlensing of cosmological quasars. Paczy\'nski (1986)
returned to the Galaxy and pointed out that
microlensing in the local group can be used to detect
or rule out astrophysical dark matter candidates.
Griest (1991) invented the popular
word MACHOs, which stands for MAssive Compact Halo Objects. 
Shortly afterwards, three groups
started microlensing surveys. Almost simultaneously, these groups
announced the discovery of microlensing events in 1993
(Alcock et al. 1993; Aubourg 1993; Udalski et al. 1993).
Another milestone in microlensing, occurred in 1994-1995,
is the capability to detect microlensing events in real-time
by the OGLE and MACHO collaborations 
(Udalski et al. 1994a; Alcock et al. 1996);
EROS now also has this
capability. Nowadays, close to ten groups are conducting microlensing
surveys toward different targets; more
than 500 microlensing events have been discovered, most of which have
been identified in real-time (\S 4).
Below we shall first review the basic concepts in
microlensing.

\section{Basic Theoretical Concepts}

The basic lensing geometry is illustrated in Fig. 1, as viewed 
sideways from the line of sight. A light
ray from a distant source is deflected by an intervening lens and
eventually reaches the observer.  For a point lens,
it is easy to show that the image positions have
to satisfy the lens equation:
\begin{equation} \label{lens}
\frac{D_{\rm d}}{D_{\rm s}} \vec{r_{\rm s}}=\vec{r}+D \vec{\alpha},~~
D \equiv \frac{D_{\rm d} D_{\rm ds}}{D_{\rm s}}, ~~
~~~\vec\alpha = -\frac{4 GM}{c^2 r} {\vec{r} \over r},
\end{equation}
where $D_{\rm d}, D_{\rm s}, D_{\rm ds}$ are the three Euclidean
distance measures (cf.
Fig. 1), $M$ the lens mass, $\alpha$ the deflection angle (twice the
Newtonian value),
and $r$ and $r_{\rm s}$ are the (physical) distances of the image and the
source from the observer-lens line, respectively. 

\begin{figure}
{
\centering \leavevmode
\hspace{2.5cm}
\epsfxsize=0.65\textwidth \epsfbox{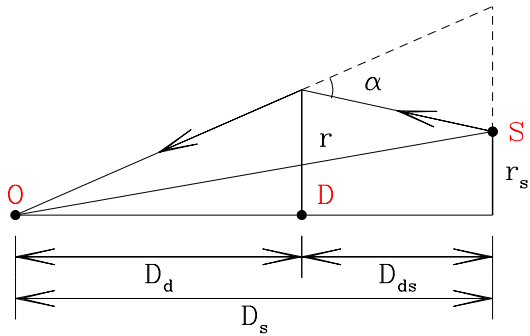}
}
\vspace{-3.8cm}
\caption{Basic geometry of gravitational lensing. `O', `D', and `S' indicate
the observer, deflector (lens) and source positions. $\alpha$ is the
deflection angle.
$r$ and $r_{\rm s}$ are the (physical) distances of the light ray and the
source from the observer-lens line, respectively.
}
\end{figure}

Eq. (\ref{lens}) can be easily solved and one finds that
there are always two images for given source position,
one with positive parity and one with negative parity 
The total magnification 
satisfies a simple relation (e.g., Paczy\'nski 1986):
\begin{equation}
\mu= \frac{u^2+2}{u\sqrt{u^2+4}},
\end{equation}
where $u=r/r_{\rm E}$. The Einstein radius $r_{\rm E}$ is given by
\begin{equation}
r_{\rm E} = \left({4GMD \over c^2}\right)^{1/2}
\approx 9 {\rm AU}
{\left({M \over M_\odot}\right)}^{1/2}
{\left({D \over 10 {\rm kpc}}\right)}^{1/2}.
\end{equation}
The angle extended by the Einstein ring on the sky is
about $\sim 1$ milli-arcsecond. Notice that when $u=1$, the
total magnification
$
\mu=3/\sqrt{5}\approx 1.342.
$
Such a change of 0.32 mag can be readily detected, so the area covered
by the Einstein ring is commonly used as a measure of the lensing
cross section.

Since everything moves in nature, the magnification changes with time due to
the change in relative alignment. So in practice we
detect microlensing by the source light change.
Two sample source trajectories and corresponding
light curves are shown in Fig. 2. The time scale is determined by
the Einstein radius crossing time
\begin{equation}
t_{\rm E} = 78 \,{\rm day}
{\left({M \over M_\odot}\right)}^{1/2}
{\left({D \over 10 {\rm kpc}}\right)}^{1/2}
{\left({V_{\rm t} \over 200 {\rm km\,s^{-1}}}\right)}^{-1},
\end{equation}
where ${V_{\rm t}}$ is the lens transverse velocity relative to the
observer-source line, and $D$ is given in eq. (\ref{lens}).
From an observed light curve, we can infer (primarily) four parameters, the
baseline magnitude of the lensed object, peak magnification, peak time
and the Einstein radius crossing time, $t_{\rm E}$. Since only the
variable $t_{\rm E}$
carries information about the lens and since
it depends on a number of parameters ($M, D_{\rm d}, D_{\rm s}$
and $V_{\rm t}$), 
it is impossible to determine the lens mass and distance uniquely. This lens
{\it degeneracy} is the most severe limitation in the current surveys and
it makes a number of interpretations ambiguous (see \S 5).
As we can see,
the `standard' light curves are {\it symmetric,
achromatic}, and {\it non-repeating} (see eq. \ref{tau}). These signatures can be used to tell
microlensing from other stellar variabilities. However,
there are exceptions to these rules, see Fig. 3 and \S 4.

So far we have concentrated on individual microlensing light curves, and
we obviously need a statistical description of the lensing sample as
a whole. For this purpose, two concepts come in handy: the optical depth
and event rate. The optical depth is the probability that
there is a lens located inside the Einstein radius at any distance along
the line of sight, or equivalently,
the probability that a source is magnified by more than
a factor of $3/\sqrt{5}$ at any given time. For any lens number
density distribution, it can be calculated as
(Paczy\'nski 1986)
\begin{equation} \label{tau}
\tau = \int_0^{D_{\rm s}} n ~ (\pi r_{\rm E}^2) ~ dD_{\rm d}.
\end{equation}
In particular, for a
self-gravitating system with a constant rotation velocity $V$,
$\tau \propto V^2/c^2$ ($\sim 5\times 10^{-7}$ for the Milky Way; the
small optical depth implies that the microlensing variability is
non-repetitive.)
Notice that the optical depth
depends only on the total mass in all lenses, but is
{\it independent of the mass function of lenses}. This can be easily
seen since $n\propto \rho/m$ where as $\pi r_{\rm E}^2 \propto m$, so the product
of $n$ and $\pi r_{\rm E}^2$ does not depend on the mass of lenses. 
The optical depth
can therefore be used to infer the {\it overall} mass distribution of the
lenses.

\begin{figure}
{
\centering \leavevmode
\hspace{2cm}
\epsfxsize=0.65\textwidth \epsfbox{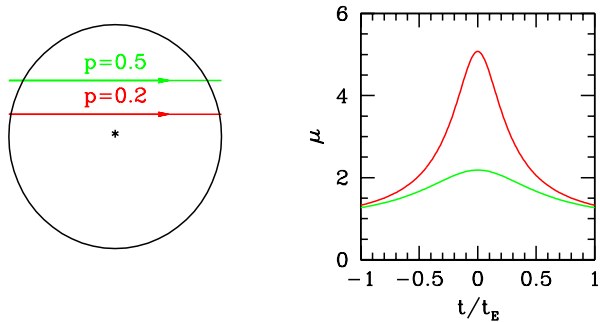} \hfil
}
\vspace{-4.25cm}
\caption{The left panel shows 
two source trajectories on the plane of sky 
with impact parameter (in units of 
$r_{\rm E}$) of $p=0.2, 0.5$, respectively. The circle indicates the
Einstein radius with the lens at the center.
The corresponding light curves are shown
in the right panel; the high amplitude one corresponds to $p=0.2$.
}
\end{figure}

The event rate describes the number of events per unit time.
It is related to the optical depth by (e.g., Paczy\'nski 1986):
\begin{equation} \label{gamma}
\Gamma \equiv \frac{dN_{\rm event}}{dt} = 
\frac{2}{\pi}~{N_{\star} ~ \tau}~\frac{1}{t_{\rm E}}~
\epsilon(t_{\rm E}).
\end{equation}
Hence, the event rate ($\Gamma$)
is proportional to the number of stars monitored ($N_\star$)
and the optical depth.
It is also inversely proportional to the
time-scale $t_{\rm E}$ (the shorter the events, the more we expect to see in
given time interval) and modulated by the detection of efficiency as a
function of $t_{\rm E}$ (nearly all microlensing surveys miss events shorter than
$\la$ one day and longer than a few years).
Notice that $\Gamma$ depends not only on the total mass in lenses
but also {\it on the mass function} (through $t_E$). Hence,
a detailed analysis of the event duration distribution
offers a unique method to determine the lens mass
function, {\it independent of light.}

\section{Current Observational Status}

The teams involved in primary microlensing surveys
are summarized in Table 1.
Most surveys are using 1-m class
telescopes combined with large-format
CCDs and large field-of-view (notable exceptions are the DUO campaign
and part of the EROS I survey which used Schmidt plates).
The last  few groups listed in Table 1 are surveying M31 -- some
promising candidates have been published, although the contamination by
variable stars (e.g., Miras) remains to be sorted out (Crotts \& Tomaney
1996; Ansari et al. 1999; for information on the MEGA experiment,
see Crotts 1999, this volume). 
These surveys use the image subtraction technique (`pixel' lensing) and have
great potentials (see \S 6).

\begin{table}
\caption{Status of Primary Microlensing Surveys}
\begin{center}
\small
\begin{tabular}{lllllll}
Team & Target(s) & Events & Time & Instrument & Field of View & Colors \\
\tableline
MACHO	& Bulge & $\sim 300$	&92-99 & 8
2Kx2K,1.3m & $42^\prime\times42^\prime$ & `R',`B' \\
 
		& LMC		&$\sim 25$ && & \\
		& SMC		&$\sim 2$	\\
\tableline
OGLE	I	& Bulge &20 & 92-95 & 2Kx2K,1m & $15^\prime
		\times15^\prime$& I (+V) \\
OGLE	II	& Bulge &$\sim 150$   & 97- & 2Kx2K,1.3m &
$15^\prime \times15^\prime$& I  (+V) \\
		& LMC &	$\sim 1$ 		& \\
		& SMC &	 	& \\
		& Sp. arms	& 1 &  \\
\tableline
EROS	I	& LMC		&$\la 2$ & 91-95& CCD+Schmidt &	&
`R',`B' \\
EROS	II	& Bulge		& $\sim 4$ &96-  & 16 2Kx2K,1m & $0.7^\circ
		\times 1.4^\circ$  & `R',`B' \\
	& LMC		& 1	&		&  \\
& SMC		& 	&	&  \\
		& Sp. arms	& 3 &  \\
\tableline
DUO		& Bulge &	13	&94-94 &1m Schmidt & &$ B_J$,R\\
\tableline
MOA		& B, L, S	& 	& 95- & 0.6m CCDs\\
\tableline
AGAPE 	& M31	& 2	& 94- & 1Kx1K,2m & $4^\prime \times4.5^\prime$&
		R, B \\
\tableline
Col/VATT& M31	&$\sim 6$& 94-98 & CCD, 1m-4m & $\sim 15^\prime \times
		15^\prime$  & R, I\\
\tableline
MEGA 	& M31	&	&	\\
\tableline	
Munich  & M31	&	&97- & 1Kx1K, 0.8m & $9^\prime \times 9^\prime$
		& R, I \\
\tableline
\end{tabular}
\end{center}
\end{table}
\normalsize

As can be seen from Table 1, $\sim 500$ hundred 
events have been observed and the number is 
still increasing day by day.
The majority of these events are discovered toward the Galactic bulge, while
about $\sim 30$ events have been seen toward LMC and SMC 
and a few toward spiral arms (Mao 1999; 
R. Ansari 1999, this volume). Perhaps about
$90\%$ of these events follow standard light curves; one example is
shown in the left panel of Fig. 3. The remaining fraction shows 
deviations and are often called ``exotic'' events. (It is, however,
unclear at the
present stage whether the procedures used to select standard events
are biased against these exotic events.) The so-called parallax
events exhibit deviations induced by the Earth motion around the Sun
(Refsdal 1966; Gould 1992);
they preferentially occur in long events where the Earth moves a substantial
distance around the Sun. A few such events have been reported
(Alcock et al. 1995; Mao 1999). There are also light curves that are
significantly modified by the finite size of the lensed stars
(Gould 1994; Witt \& Mao 1994; Nemiroff \& Wickramasinghe 1994); $\sim 2$ such
events have been reported (Alcock et al. 1997b).
The most dramatic deviation occurs when the lens is a
binary (e.g., Mao \& Paczy\'nski 1991).
In some cases, the light curves can show multiple peaks with
sharp rises and falls. One example is shown in the right panel of
Fig. 3, taken from Rhie (1999; see also Afonso et al. 1999). About 30
such binary microlensing events
have been seen (e.g., Udalski et al. 1994b; Alcock et al. 1999d).
There are also other deviations, such as those due to
binary sources (Griest \& Hu 1993) and wide binaries
(Di Stefano \& Mao 1996). All these deviations have been predicted
and later seen in the experiments. The importance of these exotic events
is that they provide additional constraints on the lensing
configuration. For example, the finite source size event and the binary
caustic crossing events can be used to determine the transverse velocity
(e.g., for 98-SMC-1, Afonso et al. 1999) while the parallax events
provide a measure of the Einstein radius projected on the observer plane
(in units of AU).

\begin{figure}
\centering \leavevmode
\epsfxsize=0.45\textwidth \epsfbox{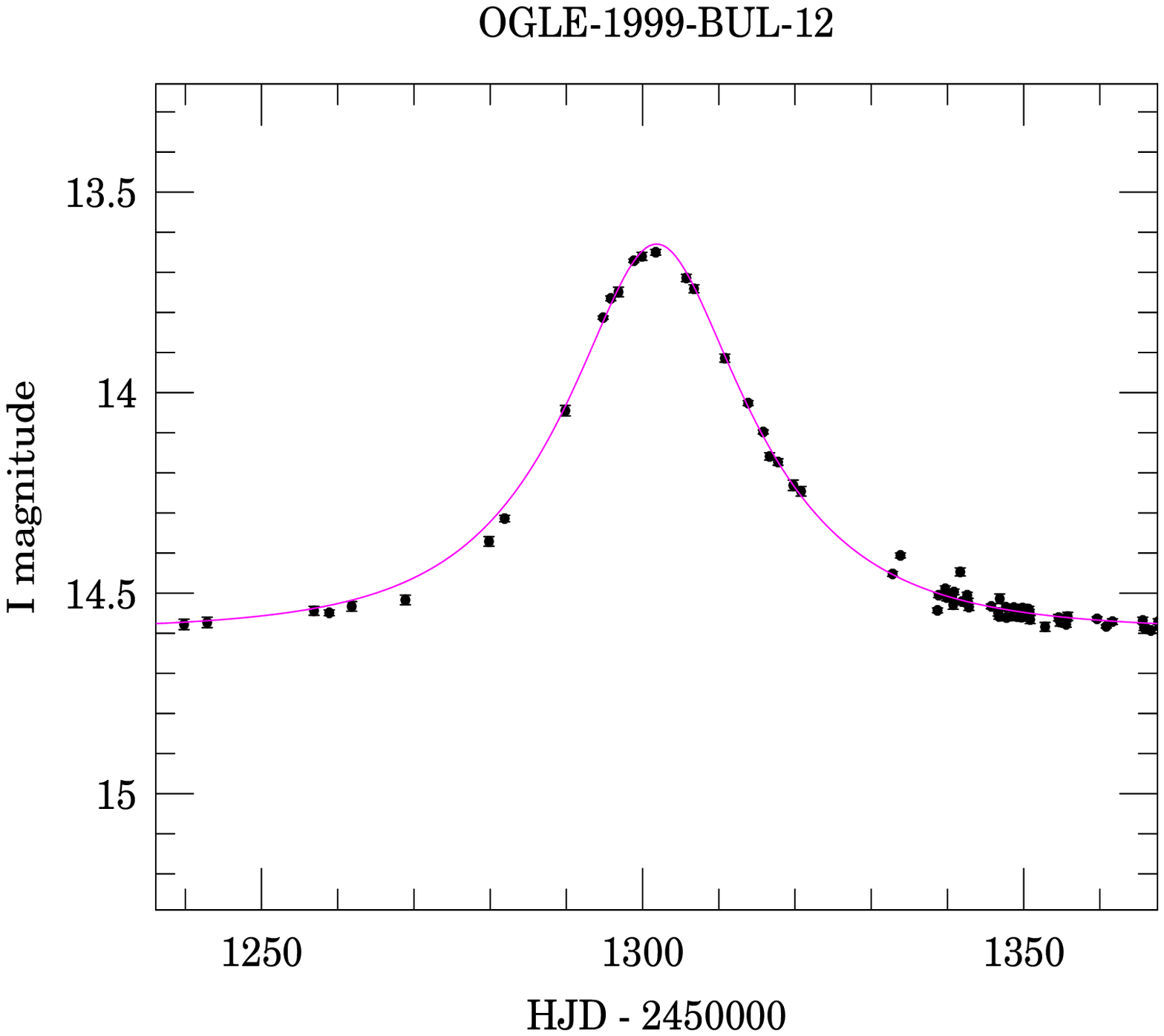}
\epsfxsize=0.41\textwidth \epsfbox{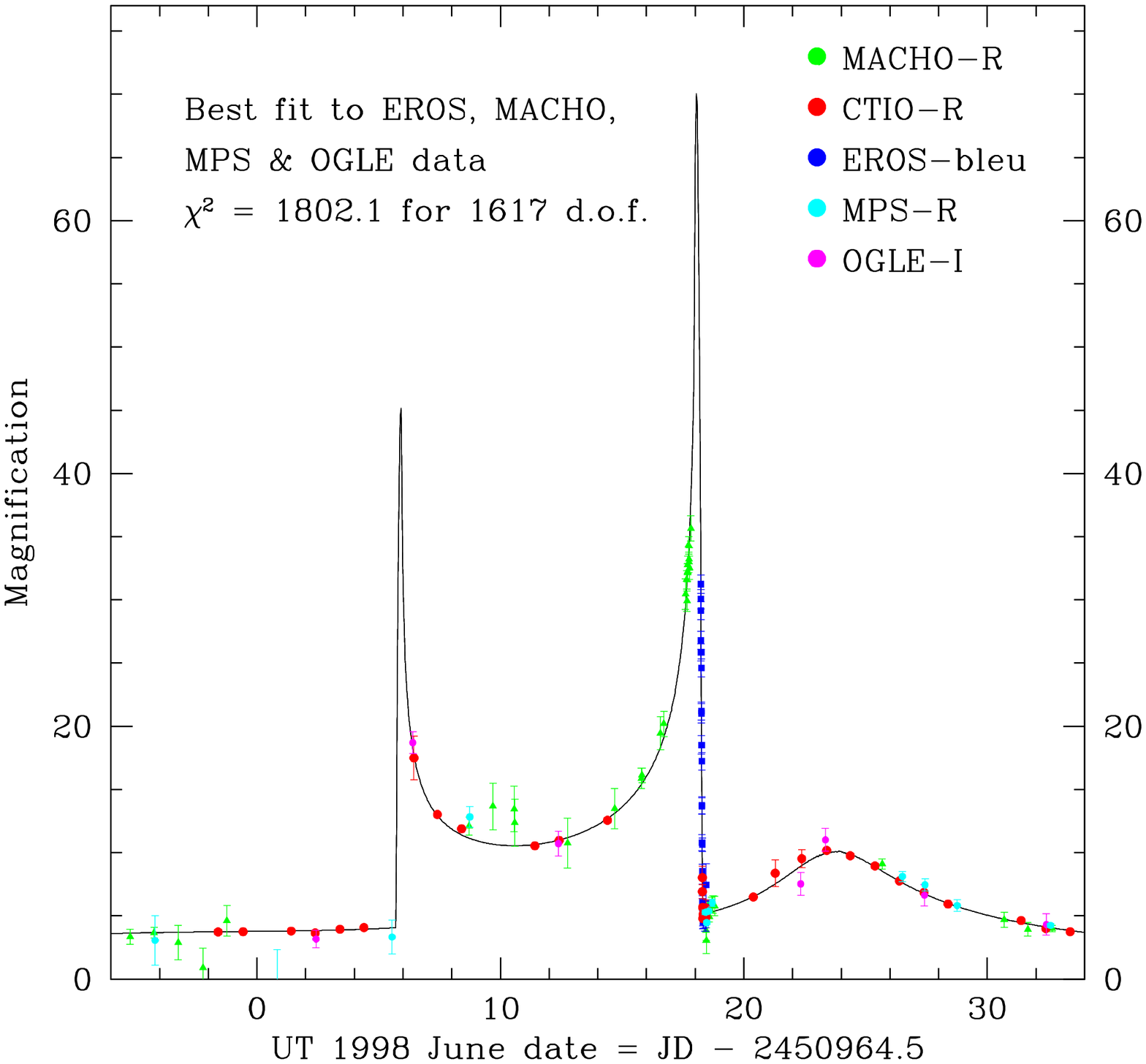}
\caption{Examples of microlensing light curves. The left panel shows a
microlensing light curve toward the Galactic bulge discovered by the
OGLE collaboration in real-time;
most events follow the symmetric, achromatic shape as
the one shown here. The right panel shows a binary
microlensing events observed by many groups (cf. Afonso et al. 1999).
}
\end{figure}

\section{Scientific Highlights}

In this section, I highlight some of the scientific returns from
the surveys. Notice that
the published microlensing results are still largely
based on $\sim 50$ events toward the bulge and $\sim 10$ events toward LMC.

\subsection{Dark Matter}

\subsubsection{Planetary Mass}

From eq. (\ref{gamma}), we see that the number of events scales with the
lens mass as $\propto M^{-1/2}$. Thus 
if the halo is dominated by low mass objects, we would
expect many short duration events. Both the EROS and MACHO collaborations
have searched for such short events toward LMC, none was found. A joint
analysis by these collaborations (shown in Fig. 4) rules out a halo
dominated by objects in the mass range of $10^{-7} M_\odot < M 
\la 10^{-2} M_\odot$
(Alcock et al. 1998).

\begin{figure}
{
\centering \leavevmode
\hspace{3cm}
\epsfxsize=0.5\textwidth \epsfbox{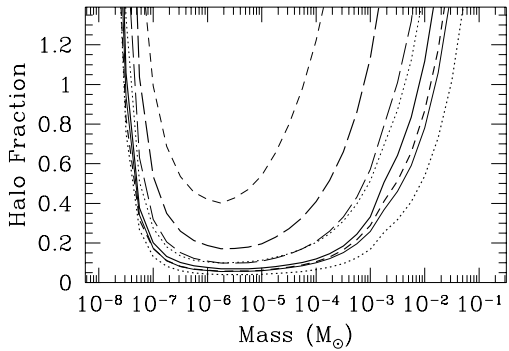}
}
\caption{Limits on the planetary dark matter objects in the Galactic
halo. Shown here are the 95\% confidence contours in the plane of the lens
mass and the fraction of mass in such lenses. The various lines
are for different halo models
with different halo flattening and rotation curves 
(adapted from Alcock et al. 1998).}
\end{figure}

\subsubsection{Dark matter with $M \sim M_\odot$}

The MACHO collaboration has analyzed $\sim 8$ microlensing
events toward LMC from their first two year experiment. They concluded
that the optical depth is $\tau \approx 2.9^{+1.4}_{-0.9} \times
10^{-7}$ and from the even durations they estimate that
the most likely lens mass is $M \sim 0.5 M_\odot$. Since
a halo full of compact objects would have $\tau \approx 5 \times
10^{-7}$, their result implies that the fraction of mass in MACHOs is 
$\sim 50\%$. Fig. 5 shows graphically these results
from a likelihood analysis (Alcock et al. 1997a).

\begin{figure}
{
\centering \leavevmode
\hspace{2.5cm}
\epsfxsize=0.6\textwidth \epsfbox{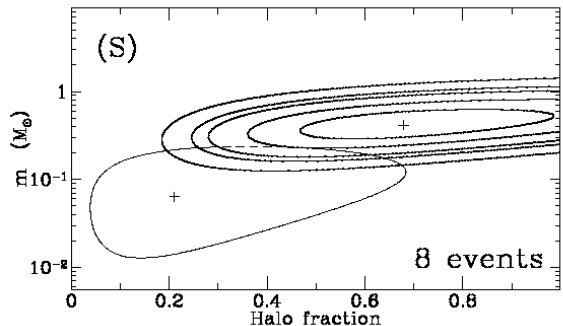}
}
\caption{
Limits on the dark matter objects in the Galactic
halo. Shown here is the likelihood contours in the plane of the lens
mass and the fraction of mass in such lenses. The contour levels are
respectively 34\%, 68\%, 90\%, 95\%, and 99\%. The lower left contour is
the 90\% limit from the first year data (adapted from Alcock et
al. 1997b).
}
\end{figure}

To understand this important result, we have to recall
the optical depths contributed by known
populations. Galactic disk populations 
contribute about $\tau \sim 1.5\times 10^{-8}$
(cf. Table 10 in Alcock et al. 1997b). The simplest estimates of $\tau$ for
self-lensing of LMC stars by LMC stars are
$\tau \approx 0.1-0.5 \times 10^{-7}$ (Wu 1994; Sahu 1994). The inferred
optical depth (which still has {\it large} uncertainties) is 
larger than these stellar contributions. So the most straightforward 
interpretation is that one has discovered an entirely new MACHO population.
Many candidates have been proposed, such as
white dwarfs, primordial black holes, dense molecular clouds etc.
All these scenarios have some difficulties. For example, if the
MACHOs are white dwarfs, these stars will produce too much
chemical enrichment in the halo (e.g., Freese et
al. 1999, and references therein).
Also the high MACHO fraction is somewhat larger than the baryon fraction
expected from nucleosynthesis. 
Nucleosynthesis predicts the density parameter in baryons
$\Omega_B \approx
0.015h^{-2}$ (Walker et al. 1991),
and hence the global baryon fraction is
\begin{equation}
f_{\rm B}
=\frac{\Omega_B}{\Omega_0}\approx 12\% \left(\frac{H_0}{65 {\rm\, km\,s^{-1}}}
\right)^{-2}
\left(\frac{\Omega_0}{0.3}\right)^{-1},
\end{equation}
where $\Omega_0$ is the total matter density and $H_0$ is the Hubble constant.
The fraction is around $\sim 10\%$ for $\Omega_0=0.3$, while
for $\Omega_0=1$, the fraction is only around 4\%. 
So there seems to be some contradictions.

In light of these difficulties, people have put forward revised 
scenarios of self-lensing, such as self-lensing of tidal LMC disk or
lensing due to intervening stellar populations along the line of sight
(Zhao 1998). The three binary microlensing events toward LMC and SMC are
consistent with self-lensing. It remains controversial whether
the stellar population studies support this scenario (see Zhao 1999, and
references therein). It is desirable to
to test different  scenarios empirically. One important
possibility is to use exotic microlensing events to obtain further
constraints on the lens locations (see Evans 1999, this volume). We can also
examine statistical properties of the lensed population
(Stubbs 1998; Zhao 1999). For example, we can examine the optical depth
across the LMC as a function of stellar density. For 
self-lensing, we expect $\tau \propto \rho_*^2$,
while for lensing by dark matter objects, $\tau \propto \rho_*$. 
Also in the self-lensing scenario, the stars at the far-side of LMC
is more likely to be lensed than those at the near-side, so the lensed sources
should be systematically fainter and suffer more extinction and
reddening (Zhao 1999 and references therein). Further, they may also have different kinematical
behaviors. We return to some future possibilities in \S 6.

\subsection{Galactic Bulge}

\subsubsection{Galactic Structure and Optical Depth}

The OGLE collaboration gives 
an optical depth
$\tau=(3.3\pm 1.2) \times 10^{-6}$
based on $\sim 10$ events (Udalski et al. 1994a), while the 
MACHO collaboration gives 
$\tau=3.9^{+1.8}_{-1.2} \times 10^{-6}$ 
(Alcock et al. 1997c) based on 45 events.

Earlier estimates showed that (Paczy\'nski 1991; Griest et al. 1991):
$\tau \sim 4-8 \times 10^{-7}$, much smaller than the observed one. 
Taken into account the self-lensing 
(Kiraga \& Paczy\'nski 1994) and galactic bar structure
(Zhao et al. 1996), $\tau$ was revised upward to
$\sim 2\times 10^{-6}$. However, a more recent dynamical model of
the Galaxy gives an optical depth of only
$1.4 \times 10^{-6}$ (H\"afner et al. 1999), which seems to be too low.
It is unclear whether the discrepancy is serious or
not, since
the observed $\tau$ is based on only $\sim 50$ events so far. Furthermore,
the effect of blending on $\tau$ has not been estimated  realistically.
If the discrepancy remains when the larger (existing) database of
events is analyzed, this signals that our dynamical
understanding of the Galaxy may be incomplete.

\subsubsection{Mass function}

Fig. 6 shows the predicted duration distribution from Zhao et al.
(1996). A Salpeter mass function  ($dN/dM \propto M^{-2.35}$)
matches the observed duration
distribution very well, while one dominated by brown dwarfs predict
too many short events and hence is ruled out. In contrast, the
predicted distribution with a disk mass function derived from
HST star counts (Han \& Gould 1996) would predict
too few short events. This is easy to understand since the disk mass function
($dN/dM \propto M^{-0.5}$) 
is significantly flatter than the Salpeter mass function at the low mass end.
The bulge mass function determined using HST NICMOS data (Zoccali et
al. 1999) may also be too flat to explain the observed distribution.
The observations hint that the mass functions at different
places in the Galaxy (in particular the disk and bulge) may be
different. Some of the differences may be due to evolutionary effects,
as evidenced in globular clusters (see Zoccali et al. 1999 and references
therein). 

\begin{figure}
{
\centering \leavevmode
\hspace{2.5cm}
\epsfxsize=0.65\textwidth \epsfbox{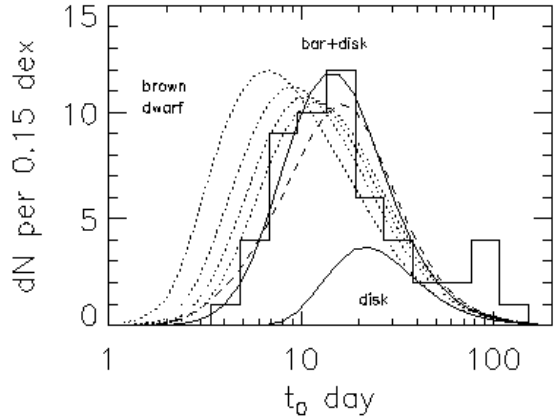}
}
\vspace{-0.5cm}
\caption{
The histogram shows the observed distribution of event durations.
The two solid lines are the predictions
for the disk population and the whole population (bar+disk). 
The dotted lines are for different
brown dwarf contributions (the leftmost one is
with a 60\% brown dwarf contribution).
See Zhao et al. (1996) for more details.
}
\end{figure}

\subsection{Stellar Studies}

The contribution of microlensing surveys to stellar studies is
multi-fold. First, it provides
excellent color-magnitude diagrams. Combined
with stellar population synthesis models, they can be used to decipher
different components in the Galaxy (e.g., Ng et al. 1996).
Second, microlensing surveys provides tens of thousands
variable stars, including cepheid variables and RR Lyrae stars.
The huge database allows one, for example,
to investigate the metallicity effects in 
cepheids, which has important implications on the 
cosmological distance ladder (Sasselov et al. 1997). These variable
stars has also been used to infer the star formation history of the LMC 
(Alcock et al. 1999a). The
rare detached eclipsing binaries can be used to
determine the distances in a single step,
to a few percent accuracy (Paczy\'nski 1996a). 
The red clump stars have also been argued as a standard candle 
(e.g., Udalski 1998) and hence
can be used to determine
distances and obtain extinction maps toward the Galactic center (Stanek 1996).
Highly magnified events are particularly suitable for real-time
spectroscopies; a pilot program has been carried out by
Lennon et al. (1996, 1997); these spectra have been used to derive the
ages, abundances of the lensed stars.
Interested readers are referred to the recent conference held in
Budapest for more information at
{\tt http://www.konkoly.hu:80/iau176/.}

\subsection{Planet Detection}

Planet detection has become an important application of 
microlensing. The method was proposed a number of years ago
(Mao \& Paczy\'nski 1991; Gould \& Loeb 1992; Bennett \& Rhie 1996; Di
Stefano \& Scalzo 1999). This field is maturing due to
the efforts by the PLANET, MPS and other collaborations.
There has been even a claim that a Jovian planet
was discovered in a binary star system (Bennett et al. 1999).
More complete references and discussions can be found in the
review by P. Sackett (this volume).

\section{Future Directions}

As discussed in \S 5, Microlensing surveys have been very successful,
producing exciting results on many branches of astronomy.
However, the current
surveys also have limitations. First,
only $\sim 30$ events have been discovered so far toward LMC and SMC
($\sim 10$ analyzed so far!), so the
results are still dominated by small number statistics. The 
lines of sight probed are still largely limited to the bulge,
LMC, SMC and M31. Most frustratingly, due to the lens degeneracy,
the lens mass, distance and hence the interpretations remain uncertain.

The future directions of microlensing research are to overcome
these weaknesses in the next few years.
{\it First}, there is still a large un-analyzed database, particularly
toward the bulge. Analyses of these events will map
the optical depth as a function of latitude and longitude and
hence probe the structure of the galaxy. The
distribution of event durations also places unique
constraints on the mass function of stars when
hundreds of microlensing are available (Mao \& Paczy\'nski 1996). For 
example, from microlensing toward LMC, SMC and the Galactic bulge it will
be possible eventually to study the metallicity dependence of the
(initial) mass functions.
The {\it second} promising direction
is the large-scale application of the image subtraction method 
(Tomaney \& Crotts 1996; Alard \& Lupton 1998; Alcock et al. 1999b,c).
The main difficulty is to take into account the
point-spread-function variations.
Apparently this difficulty can be overcome, particularly with the
non-constant kernel smoothing (Alard 1999).
The method seems to provide more precise photometries, 
as a result,
the event number is increased by a factor of $\sim 2$ toward the bulge.
So far this method has been applied to limited data set,
wider applications of this method will be an important advance
on the software level.
{\it Thirdly}, although the MACHO collaboration will end
in 1999, the EROS II and OGLE II experiments
will continue their observations.
The OGLE II experiment plans to install a larger camera.
This will allow the detection of a larger number of microlensing
events. Another proposed experiment (Stubbs 1998) will increase the
number of events toward LMC by orders of magnitude.
This will put the microlensing results toward LMC on a sound 
statistical footing. {\it Fourthly},
the Space Interferometry Mission
(to be launched in 2005), with astrometric
accuracy of $\sim$ few micro-arcsecond, will in principle 
allow one to determine lens mass and distance uniquely for a number of
events, and hence removing the lens degeneracy 
(e.g., Paczy\'nski 1998; Boden, Shao \& Van Buren 1998).
For the followup studies, continued efforts
to detect planets using microlensing should be very
fruitful. Also a more systematic spectroscopic survey 
on 8-10m class telescopes will yield reliable
ages and chemical abundances for dwarfs in the bulge;
such information will be important for
understanding the formation and evolution of the Galaxy.

\acknowledgments
We are very grateful to Z. Zheng and W. Lin for 
helpful comments on a draft of this manuscript.

\end{document}